# Phase diagram and superconducting dome of infinite-layer $Nd_{1-x}Sr_xNiO_2$ thin films

Shengwei Zeng,[1,2,*] Chi Sin Tang,[1,3,4] Xinmao Yin,[1,4] Changjian Li,[2,5] Mengsha Li,[5] Zhen Huang,[2] Junxiong Hu,[1,2] Wei Liu,[1] Ganesh Ji Omar,[1,2] Hariom Jani,[2,3] Zhi Shiuh Lim,[1,2] Kun Han,[2] Dongyang Wan,[1,2] Ping Yang,[4] Stephen John Pennycook[2,5], Andrew T. S. Wee,[1,3,4,6] Ariando Ariando,[1,2,3,6,*]

[1]Department of Physics, Faculty of Science, National University of Singapore, Singapore 117551, Singapore
[2]NUSNNI-NanoCore, National University of Singapore, Singapore 117411, Singapore
[3]NUS Graduate School for Integrative Sciences and Engineering, National University of Singapore, Singapore 117456, Singapore
[4]Singapore Synchrotron Light Source (SSLS), National University of Singapore, Singapore 117603, Singapore
[5]Department of Materials Science and Engineering, National University of Singapore, Singapore 117575, Singapore
[6]Centre for Advanced 2D Materials and Graphene Research, National University of Singapore, Singapore 117546, Singapore
*To whom correspondence should be addressed.
E-mail: phyzen@nus.edu.sg, ariando@nus.edu.sg

Infinite-layer $Nd_{1-x}Sr_xNiO_2$ thin films with Sr doping level $x$ from 0.08 to 0.3 were synthesized and investigated. We found a superconducting (SC) dome between $0.12 < x < 0.235$ accompanied by a weakly insulating behaviour in both underdoped and overdoped regimes. The dome is akin to that in the electron-doped 214-type and infinite-layer cuprate superconductors. For $x \geq 0.18$, the normal state Hall coefficient ($R_H$) changes the sign from negative to positive as the temperature decreases. The temperature of the sign-changes decreases monotonically with decreasing $x$ from the overdoped side and approaches the superconducting dome at the mid-point, suggesting a reconstruction of the Fermi surface with the dopant concentration across the dome.

The recent discovery of superconductivity in doped infinite-layer nickelate $Nd_{0.8}Sr_{0.2}NiO_2$ thin films has hailed another milestone in the field of high-temperature superconductivity [1]. In contrast to the antiferromagnetic (AF) order in the cuprates, the parent compound $NdNiO_2$ shows no sign of magnetic order down to a temperature of 1.7 K [2], despite its similarities in the crystalline and electronic structures to the high-$T_c$ cuprates. This severely challenges the scenario of AF spin fluctuation as the superconducting pairing mechanism, which is prevalent in the cuprates [3-5]. The nickelate may represent a new type of superconductivity and has motivated numerous theoretical work to explore the similarities and differences to the cuprates [6-25]. Among the similarities are the large value of the $R$NiO$_2$ ($R$ = La, Nd) long-range hopping $t'$ and the $e_g$ energy splitting [6] and the possible $d$-wave pairing symmetry as suggested by the $t$-$J$ model [12, 13]. It has been proposed however that the coupling between the low-density Nd-$5d$ conduction electrons and the localized Ni-$3d$ electron forms Kondo spin singlets and thus suppresses the long-range AF order [15]. Further, while the doped holes in the $Nd_{1-x}Sr_xNiO_2$ reside at the Ni sites, they are at the oxygen sites in the cuprates due to the strong hybridization of the Cu-$3d_{x2-y2}$ and O-$2p$ orbitals [11, 14]. The role of Nd $4f$ state in $NdNiO_2$, which shows no substantial effect in cuprates, has also been considered [7, 10].

Experimentally, X-ray spectroscopy suggests a weakly interacting three-dimensional $5d$ metallic state in the rare-earth spacer layer hybridizing with the strongly correlated $3d_{x2-y2}$ state in the $NiO_2$ layer [26]. Spin fluctuations in bulk $NdNiO_2$ have been explored by Raman spectroscopy which suggested an AF exchange of $J$ = 25 meV [27]. $Nd_{1-x}Sr_xNiO_2$ ($x$ = 0, 0.2, 0.4) prepared in bulk form shows the absence of superconductivity even under 50.2 GPa pressure, probably due to the creation of nickel deficiency during the reduction process [28]. Besides, attempts in growing the nickelate infinite-layer films at low oxygen partial pressure without chemical reduction have resulted in insulating thin films [29]. It is clear that the synthesis of the superconducting nickelate is challenging and critical.

One important feature in most unconventional superconductors is that the superconductivity is induced by charge doping in the parent compound. Variation in the dopant concentration leads to a phase diagram in which a SC dome is embedded within other competing ordered phases such as insulator and metal [30, 31]. For the isostructural cuprates, increasing charge doping causes transitions from a Mott insulating to a high-$T_c$ superconducting and then a Fermi-liquid metallic state [30]. It is important to note here that the use of the Nd atom and the $x = 0.2$ value in the $Nd_{0.8}Sr_{0.2}NiO_2$ superconductor has been inspired by the need to achieve a higher conductivity in the $La_{0.8}Sr_{0.2}NiO_2$ through substituting the La with a smaller ionic radius atom Nd and thus increasing the electronic bandwidth [1]. In contrary to the expectation of hole doping by Sr, the normal-state $R_H$ of the $Nd_{0.8}Sr_{0.2}NiO_2$ showed a negative sign at room temperature. However, whether the superconducting and normal state properties in $Nd_{1-x}Sr_xNiO_2$ are doping-dependent thus far remains unresolved [32]. In this letter, we report the synthesis of the infinite-layer $Nd_{1-x}Sr_xNiO_2$ with different Sr doping levels showing a SC dome for $0.12 < x < 0.235$ with a weak insulating state in both the underdoped and overdoped regimes beside the dome. The overdoped regime behaves differently from the high-$T_c$ cuprates, wherein Fermi liquid metal behaviour is seen.

The ceramic targets were prepared by sintering the mix of the $Nd_2O_3$, $SrCO_3$ and NiO powder (weighted according to the chemical formula of $Nd_{1-x}Sr_xNiO_3$) in the air for 15 h at 1200, 1220 and 1250 °C, respectively, with a regrinding process before each sintering. For the final sintering at 1250 °C, the powder was pressed into a disk-shaped pellet. Thin films with a thickness of ~35 nm (unless otherwise indicated) were grown on a $TiO_2$-terminated (001) $SrTiO_3$ (STO) substrate using a pulsed laser deposition (PLD) technique with a 248-nm KrF excimer laser. The nominal composition of the as-grown thin films is assumed to be the same as the target. No capping layer is introduced for all samples in this study. The deposition temperature and oxygen partial pressure $P_{O2}$ for all samples were 600 °C and 150 mTorr, respectively. The laser energy density on the target surface was set to be 2 Jcm$^{-2}$. After deposition, the samples were cooled down to room temperature at a rate of 8 °C/min at deposition pressure. It has been suggested that to induce a superconducting phase, the infinite-layer nickelate needs to be chemically reduced by annealing the sample and $CaH_2$ powder in a vacuum that stimulates the $H_2$ gas-phase reaction with the sample [1, 2, 33-35]. In our case, we performed this chemical reduction in the PLD vacuum chamber after cutting the samples into pieces of size around 1.3 × 2.5 mm$^2$ and wrapping them in an aluminium foil together with about 0.1 g of $CaH_2$ powder (the samples were embedded in $CaH_2$ powder). During the reduction process, the samples were heated up to 340-360 °C at a rate of 25 °C/min, kept at that temperature for 80-120 mins, and then cooled down to room temperature at a rate of 25 °C/min. We note here that the pressure changed from the background of 1×10$^{-5}$ Torr to around 1×10$^{-1}$ Torr during the heating, which was then maintained during the whole reduction process. The increase in pressure may be caused by the release of $H_2$ gas upon $CaH_2$ heating and/or $H_2O$ gas due to a chemical reaction between the sample and $CaH_2$ [36]. The X-ray diffraction (XRD) measurement was done in the X-ray Diffraction and Development (XDD) beamline at Singapore Synchrotron Light Source (SSLS) with an X-ray wavelength of $\lambda$ = 1.5404 Å. The wire connection for the electrical transport measurement was made by Al ultrasonic wire bonding. The transport measurements were performed using a Quantum Design Physical Property Measurement System (PPMS). The high-angle annular dark-field scanning transmission electron microscopy (HAADF-STEM) imaging was carried out at 200 kV using a JEOL ARM200F microscope and the cross-sectional TEM specimens were prepared by a focused ion beam machine (FEI Versa 3D).

Figure 1 depicts the XRD $\theta$–$2\theta$ patterns of the $Nd_{1-x}Sr_xNiO_2$ thin films with a Sr doping level $x$ from 0.08 to 0.3 after the chemical reduction, showing only the (00$l$) infinite layer peaks ($l$ is an integer) which is consistent with a transformation of the $Nd_{1-x}Sr_xNiO_3$ from a perovskite to an infinite-layer structure. The thin film peaks also slightly shift towards a lower angle as $x$ increases, indicating an expansion of the $c$-axis lattice constants $d$ from 3.30 Å at $x$ = 0.08 to 3.42 Å at $x$ = 0.3 as plotted in the inset of Fig. 1. This result is in agreement with the empirical expectation when replacing the cation with an atom having a larger ionic radius. The full width at half-maximum (FWHM) of the rocking curves for the (002) peaks shows a value between 0.05° and 0.11°, indicating a good quality of the infinite-layer film (Fig. S1 in the Supplementary Material) [37]. The structural characterization of the as-grown $Nd_{1-x}Sr_xNiO_3$ thin films can be found in Fig. S2 [37] for comparison.

Figure 2 shows the resistivity versus temperature ($\rho$-$T$) curves of the $Nd_{1-x}Sr_xNiO_2$ thin films for $x$ from 0.08 to 0.3. For $0.08 \leq x \leq 0.12$, the samples show a metallic behaviour at high temperatures with a resistivity minimum between 27 - 35 K, below which a weakly insulating behaviour is seen. For $0.135 \leq x \leq 0.22$, the samples are superconducting and the $\rho$-$T$ curves behave like a metal above the superconducting transition temperature. Figure 2(b) shows the zoomed-in $\rho$-$T$ curves near the superconducting transition temperatures for $0.135 \leq x \leq 0.22$. For the higher doping levels $x \geq 0.235$, the resistivity behaviour is similar to the underdoped ones but with a minimum between 12 - 18 K. The weakly insulating behaviour at low temperature in unconventional superconductors has also been observed in underdoped cuprates and attributed to the localization and/or Kondo effect [38, 39]. In nickelates, a theoretical calculation has suggested the Kondo effect to explain the resistance upturn in undoped $LaNiO_2$ and $NdNiO_2$ [15]. With Sr doping in $Nd_{1-x}Sr_xNiO_2$, both the underdoped and overdoped sides also show such a resistance upturn, the origin of which needs further investigation. To further confirm the emergence of the superconductivity, we demonstrate a suppression of the $T_c$ with magnetic fields in the $\rho$-$T$ curves for $x = 0.2$ as shown in Fig. S3 [37]. The $\rho$-$T$ curves of the as-grown $Nd_{1-x}Sr_xNiO_3$ and additional $Nd_{1-x}Sr_xNiO_2$ samples for $x = 0.12$-$0.3$ films can be found in Fig. S4 and Fig. S5 [37], respectively, signifying the reproducibility of the samples and the importance of the reduction process.

Figure 3(a) shows the temperature dependence of the normal-state $R_H$ for the $Nd_{1-x}Sr_xNiO_2$ films. At room temperature, the $R_H$ of all samples shows a negative sign and its magnitude ($|R_H|$) decreases with increasing $x$ and then saturates at $x > 0.2$ (Fig. 3(b)). This $R_H$ behaviour indicates that the charge carriers are electrons with a density increasing with $x$ (considering the carrier density $n = 1/eR_H$, where $e$ is the electron charge), in contrast to the expectation of a single-band hole doping as a result of the Nd-by-Sr substitution. While the $R_H$ remains negative below 300 K for samples with $x \leq 0.15$, it becomes positive at low temperature for samples with $x \geq 0.18$. Figure 3(b) presents the doping dependence of the $R_H$ at 20 K, clearly showing a sign-change at $x = 0.15\text{-}0.18$ from negative to positive with increasing $x$. For the samples with $x \leq 0.135$, the $|R_H|$ increases moderately with decreasing the temperature (down to 100-200 K depending on $x$) and then decreases again with further decreasing the temperature. This behaviour is similar to the observation in both electron-doped $Pr_{2-x}Ce_xCuO_4$ and hole-doped $La_{2-x}Sr_xCuO_4$ cuprates [40-42]. For samples with $x = 0.15$, the $|R_H|$ decreases monotonically as the temperature is reduced. For samples with $x \geq 0.18$, the $|R_H|$ decreases with decreasing temperature at the negative side and increases at the positive side. This indicates cancellation of the electron and hole contribution as the temperature decreases for those samples. Even though the electron density is higher (lower $|R_H|$) at a higher Sr doping level, the hole contribution is much more obvious, leading to a higher temperature where the $R_H$ changes from negative to positive sign (denoted as $T_H$) at higher $x$ (Fig. 4).

Even though the room-temperature electron density generally increases with $x$, the $\rho$ does not follow a similar trend at the overdoped regime $x \geq 0.235$ (Fig. 2). Besides, the $\rho$ and $|R_H|$ of our thin films are higher than those in previous reports [1, 32]. These could be due to the higher temperature (340 - 360 °C) of the reduction process in our experiment. Indeed, previous results showed that the resistivity of the $LaNiO_2$ increased with increasing reduction temperature even though the films showed similar infinite-layer XRD peaks [36]. The possibility of having more intercalating hydrogen at a higher reduction temperature may also influence the $\rho$ and $|R_H|$ as theoretically predicted in Ref. [43]. As the film thickness increases, moreover, the Ruddlesden–Popper-type phase can emerge as can be seen from the HAADF-STEM image (Fig. S6 in the Supplementary Material) [33, 37], which may also cause an increase in the $\rho$. Even though there exists a variation in $\rho$ due to the difference in the thickness and reduction temperature, the superconductivity and the superconducting transition temperature are reproducible and robust for different dopant levels. The resultant phase diagram is therefore conclusive.

Figure 4 depicts a phase diagram of the $Nd_{1-x}Sr_xNiO_2$ showing a SC dome between $0.12 < x < 0.235$, consistent with that in a recent report [32]. The SC dome of the $Nd_{1-x}Sr_xNiO_2$ is comparable to that of electron-doped cuprates $Sr_{1-x}La_xCuO_2$ and $Pr_{2-x}Ce_xCuO_4$ [44-46], which is much narrower compared to the hole-doped cuprate $La_{2-x}Sr_xCuO_4$ [47]. A small dip at the top of the SC dome was observed at $x = 0.18$, which is also visible in Ref. [32] but at $x = 0.2$. A similar feature has been observed in $La_{2-x}M_xCuO_4$ ($M$ = Sr and Ba) at $x = 1/8$ which was ascribed to the presence of the charge stripe [47, 48]. The $T_H$ for the samples with $x \geq 0.18$ is also shown in the phase diagram in Fig. 4. As $x$ decreases from the overdoped side, $T_H$ decreases monotonically. A simplified two-band model with both electron- and hole-like Fermi surfaces is shown schematically in the inset of Fig. 4. The $T_H$-$x$ curve separates the normal state regime into two panels with the right panel dominated by the hole-like band and the left panel by the electron-like band. With decreasing the temperature and increasing the Sr concentration, the bands dominating the normal state electrical transport evolve from an electron- to a hole-like band. Therefore, the $R_H$ changes sign from negative to positive. It is worth noting that the extension of the $T_H$-$x$ curve reaches the middle of the superconducting dome, suggesting a reconstruction of the Fermi surface as the doping changes across the middle point of the dome.

The two-band model can be captured by the Hall resistance $R_{xy}$ as a function of magnetic field $B$ [32, 49, 50]. Considering the opposite sign of the electron and hole carrier and $R_H = R_{xy}/B$, the $R_H$ in the simple two-band model can be written as:

$$R_H = \frac{1}{e} \frac{(n_h\mu_h^2 - n_e\mu_e^2) + (\mu_h\mu_e B)^2(n_h - n_e)}{(n_h\mu_h + n_e\mu_e)^2 + (\mu_h\mu_e B)^2(n_h - n_e)^2},$$

where $n_e$ and $n_h$ are the electron and hole density, $\mu_e$ and $\mu_h$ are the electron and hole mobility, respectively, and $e$ is the elementary charge [32, 49, 50]. In a low-field limit, the $R_H$ can be simplified as:

$$R_H = \frac{1}{e}\frac{n_h\mu_h^2 - n_e\mu_e^2}{(n_h\mu_h + n_e\mu_e)^2}.$$

Therefore, the difference between the $n_e$ and $n_h$ and the temperature dependencies of the $\mu_e$ and $\mu_h$ cause the $R_H$ sign changes with the doping and temperature as observed in Figs. 3 and 4. This model predicts a nonlinearity in $R_{xy}$ at high fields and a fit to $R_{xy}$ can provide an estimate to the density and mobility of both hole and electron carriers [49, 50]. However, in our measurements, the $R_{xy}$ shows a perfect linear behavior up to 9 T (Fig. S8 in the Supplementary Material) [37], which prevents us from quantitatively extracting the value of $n_e$ and $n_h$. Nevertheless, the two-band model is consistent with the picture constructed by the electronic structure calculations which show the hole pocket originating from a Ni-$3d_{x2-y2}$ orbital and the electron pockets from the rare-earth element $5d_{xy}$ and $5d_{3z2-r2}$ orbitals in the parent compound [6-8, 23-26]. Moreover, upon hole doping, the 5d electron pocket diminishes and/or the new 3d hole pocket evolves, which may explain the observed sign-change of the $R_H$ as the Sr composition increases [6, 24].

In summary, we have synthesized infinite-layer Nd$_{1-x}$Sr$_x$NiO$_2$ thin films with a doping level $x$ from 0.08 to 0.3 and found a SC dome between 0.12 < $x$ < 0.235. In both underdoped and overdoped regimes adjacent to the dome, a weakly insulating behaviour is observed, which is different from the high-$T_c$ cuprates in which a Fermi liquid metal is seen in the overdoped regime. The width of the dome is comparable to that of the electron-doped infinite-layer Sr$_{1-x}$La$_x$CuO$_2$ and 214-type Pr$_{2-x}$Ce$_x$CuO$_4$. Finally, we have demonstrated a synthesis process of the infinite-layer nickelate thin films via a chemically topotactic reduction process in a sealed vacuum chamber, facilitating future studies of an *in-situ* reaction by-product monitoring for better control of the reduction process.

**Acknowledgments**

This research is supported by the Agency for Science, Technology, and Research (A*STAR) under its Advanced Manufacturing and Engineering (AME) Individual Research Grant (IRG) (A1983c0034), the National University of Singapore (NUS) Academic Research Fund (AcRF Tier 1 Grants No. R-144-000-391-144 and No. R-144-000-403-114), and the Singapore National Research Foundation (NRF) under the Competitive Research Programs (CRP Grant No. NRF-CRP15-2015-01). Ping Yang is supported by Singapore Synchrotron Light Source (SSLS) via NUS Core Support C-380-003-003-001. The authors would also like to acknowledge the SSLS for providing the facility necessary for conducting the research. The Laboratory is a National Research Infrastructure under the National Research Foundation (NRF) Singapore.

**Figures and Captions:**

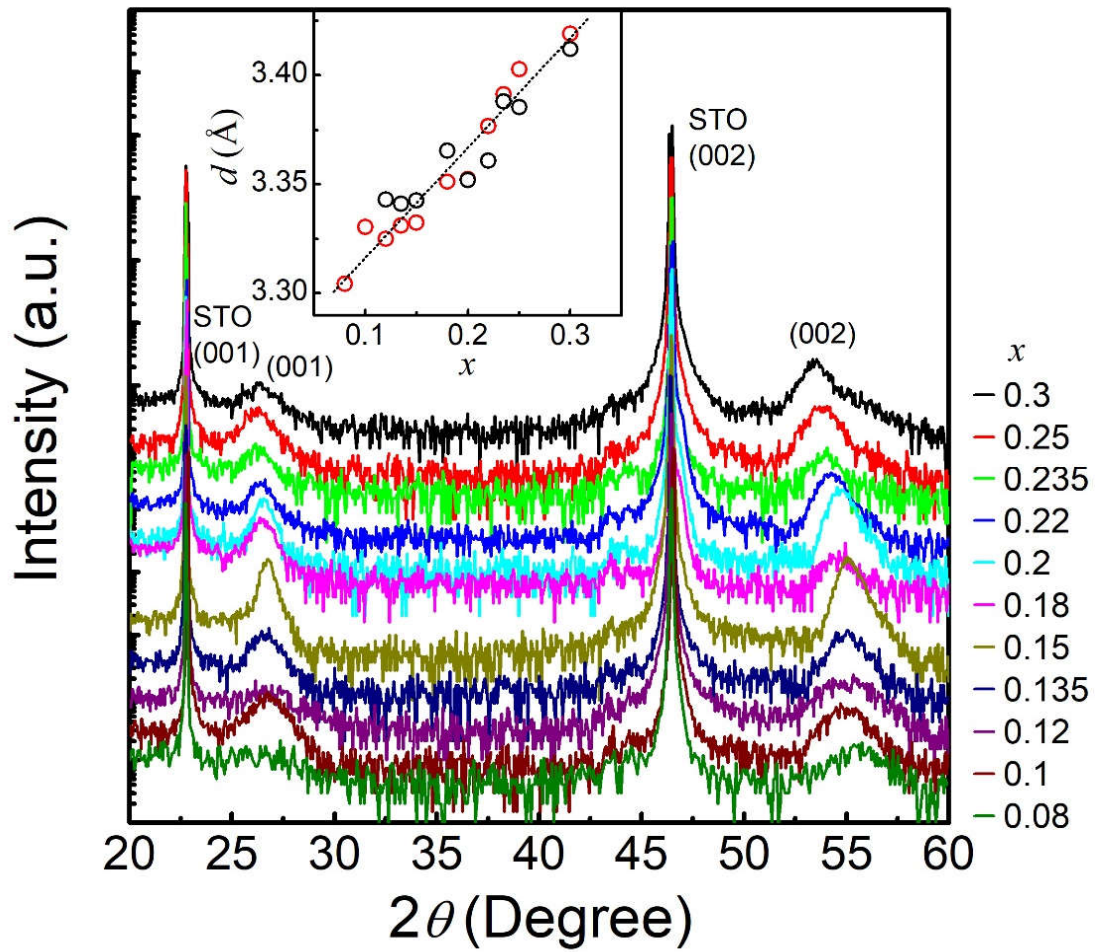

Fig. 1. The XRD $\theta$–$2\theta$ scan patterns of the $Nd_{1-x}Sr_xNiO_2$ thin films on a STO substrate. The intensity is vertically displaced for clarity. The inset shows room-temperature $c$-axis lattice constants, $d$, as a function of Sr doping level $x$. The red circles represent the data extracted from the $\theta$–$2\theta$ scans in Fig. 1 and the black circles represent another set of samples at each doping level for $0.12 \leq x \leq 0.3$. The dashed line in the inset is a guide to the eye. The measurements were done at room temperature.

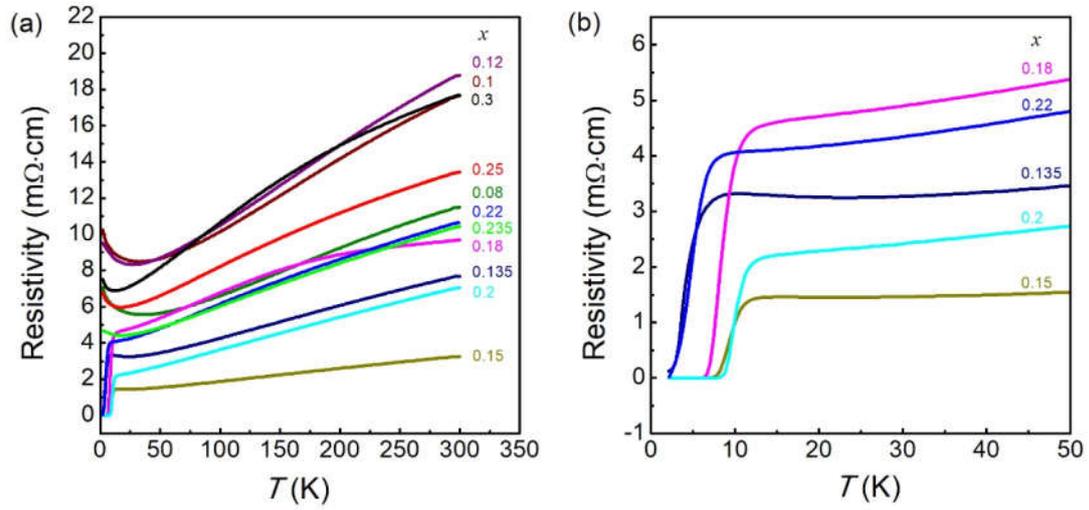

Fig. 2. (a) The resistivity versus temperature ($\rho$-$T$) curves of the $Nd_{1-x}Sr_xNiO_2$ thin films with Sr doping level $x$ from 0.08 to 0.3. (b) The zoomed-in $\rho$-$T$ curves at temperatures from 50 to 2 K for the superconducting $Nd_{1-x}Sr_xNiO_2$ with $x$ from 0.135 to 0.22.

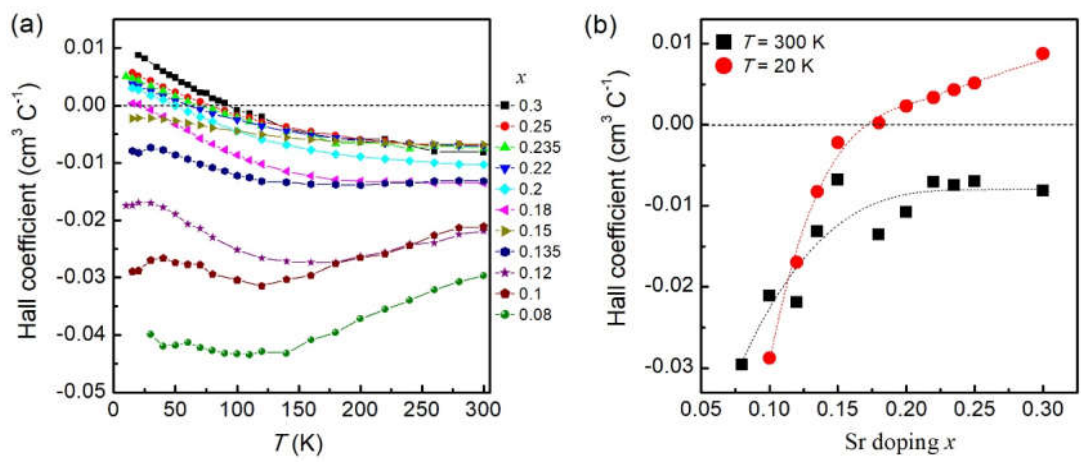

Fig. 3. (a) The temperature dependence of the $R_H$ for the $Nd_{1-x}Sr_xNiO_2$ thin films with a Sr doping level $x$ from 0.08 to 0.3. (b) The $R_H$ at $T = 300$ and 20 K as a function of $x$. The dash lines are guides to the eye.

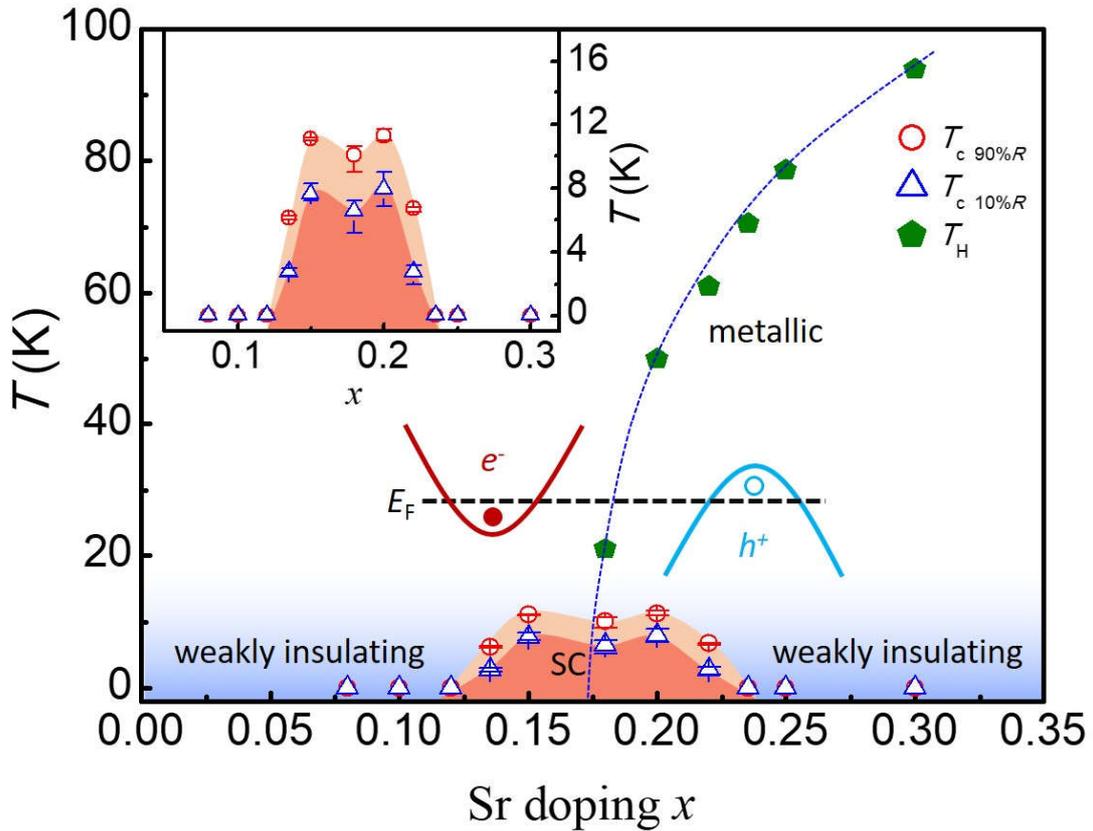

Fig. 4. The critical temperature, $T_c$, and temperature where the $R_H$ changes from negative to positive sign, $T_H$. The $T_{c\,90\%R}$ ($T_{c\,10\%R}$) are defined to be the temperature at which the resistivity drops to 90 % (10 %) of the value at 15 K (the onset of the superconductivity). The red circles and blue triangles represent the average $T_c$ of the samples shown in Fig. 2 and Fig. S5. The error bars represent the $T_c$ variation. The $T_H$ is defined to be the temperature at which the $R_H$ curve crosses the axis at zero in Fig. 3. The blue dash line is a guide to the eye. The inset in the left upper corner is the zoomed-in view of the SC dome. The inset at the bottom schematically shows the electron band ($e^-$) and hole band ($h^+$) which dominate the charge carriers in the two panels separated by the $T_H$-$x$ curve. The black dash line in the inset is the Fermi energy ($E_F$).

# Supplementary materials for

## Phase diagram and superconducting dome of infinite-layer $Nd_{1-x}Sr_xNiO_2$ thin films

Shengwei Zeng,[1,2,*] Chi Sin Tang,[1,3,4] Xinmao Yin,[1,4] Changjian Li,[2,5] Mengsha Li,[5] Zhen Huang,[2] Junxiong Hu,[1,2] Wei Liu,[1] Ganesh Ji Omar,[1,2] Hariom Jani,[2,3] Zhi Shiuh Lim,[1,2] Kun Han,[2] Dongyang Wan,[1,2] Ping Yang,[4] Stephen John Pennycook[2,5], Andrew T. S. Wee,[1,3,4,6] Ariando Ariando,[1,2,3,6,*]

[1]Department of Physics, Faculty of Science, National University of Singapore, Singapore 117551, Singapore
[2]NUSNNI-NanoCore, National University of Singapore, Singapore 117411, Singapore
[3]NUS Graduate School for Integrative Sciences and Engineering, National University of Singapore, Singapore 117456, Singapore
[4]Singapore Synchrotron Light Source (SSLS), National University of Singapore, Singapore 117603, Singapore
[5]Department of Materials Science and Engineering, National University of Singapore, Singapore 117575, Singapore
[6]Centre for Advanced 2D Materials and Graphene Research, National University of Singapore, Singapore 117546, Singapore
*To whom correspondence should be addressed.
E-mail: phyzen@nus.edu.sg, ariando@nus.edu.sg

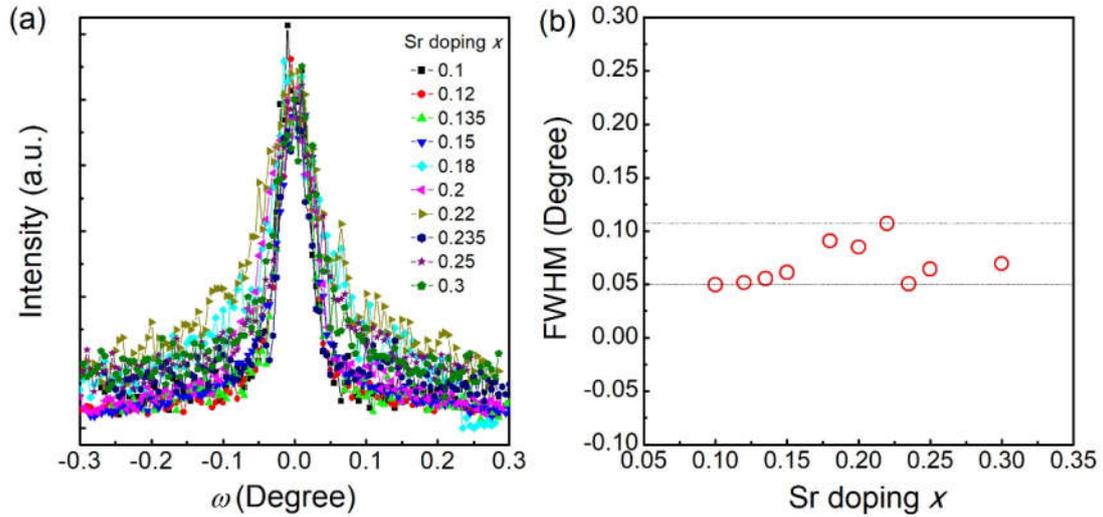

Fig. S1. (a) The rocking curves for the (002) peaks of the infinite-layer Nd$_{1-x}$Sr$_x$NiO$_2$ thin films of different $x$. (b) The full width at half-maximum (FWHM) of the (002) rocking curves as a function of $x$. The value of FWHM is between 0.05° and 0.11°.

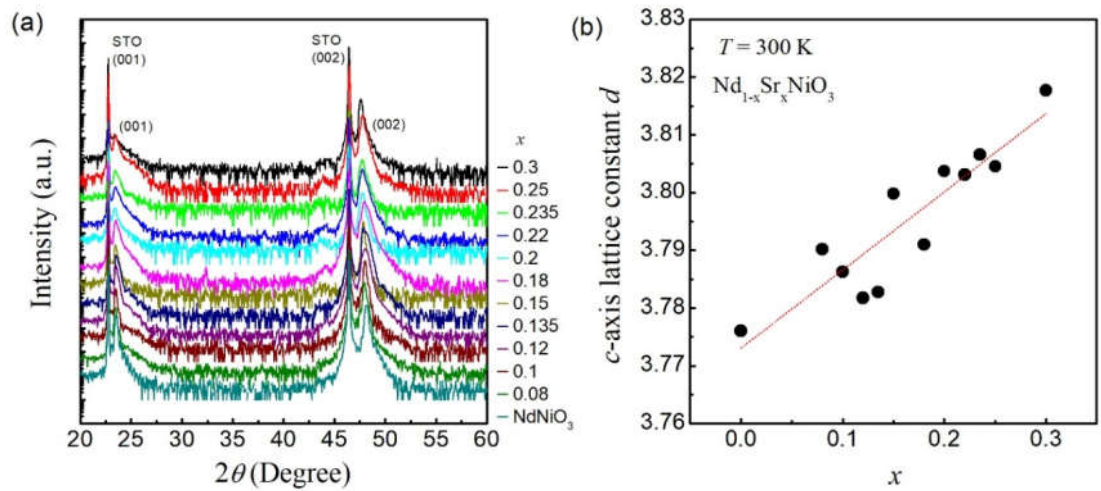

Fig. S2. (a) The XRD $\theta$–$2\theta$ scan patterns of the as-grown Nd$_{1-x}$Sr$_x$NiO$_3$ thin films on a STO substrate. The intensity is vertically displaced for clarity. Only the (00$l$) perovskite peaks are observed, where $l$ is an integer, confirming the $c$-axis oriented epitaxial growth. (b) The room-temperature $c$-axis lattice constants of the Nd$_{1-x}$Sr$_x$NiO$_3$ thin films as a function of Sr doping level $x$. The red dot line is a guide to the eye. The $c$-axis lattice constants, in general, increase slightly with $x$ from 3.776 Å for $x$ = 0 to 3.818 Å for $x$ = 0.3. The $c$-axis lattice constant obtained for the

NdNiO$_3$ ($x$ = 0) film is lower than that of the bulk (3.81 Å for pseudocubic lattice) due to the epitaxial tensile strain imposed by the STO substrate.

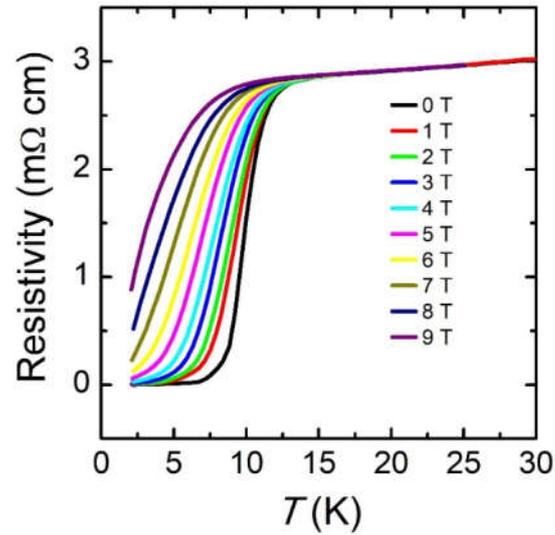

Fig. S3. The resistivity versus temperature ($\rho$-$T$) curves for the sample with $x$ = 0.2 under various magnetic field applied perpendicularly to the $a$-$b$ plane.

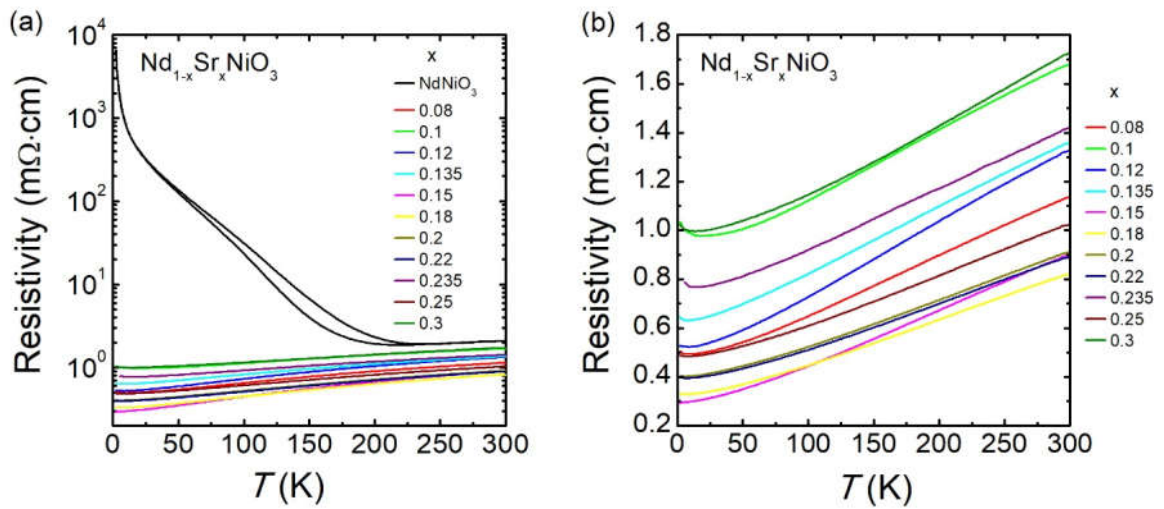

Fig. S4. (a) The logarithmic-scale resistivity versus temperature ($\rho$-$T$) curves for the undoped NdNiO$_3$ and Sr-doped Nd$_{1-x}$Sr$_x$NiO$_3$ thin films with different Sr doping level $x$ from 0.08 to 0.3. (b) The linear-scale $R$-$T$ curves for the Nd$_{1-x}$Sr$_x$NiO$_3$ thin films with $x$ from 0.08 to 0.3. The NdNiO$_3$ thin film shows a metal-insulator transition with decreasing temperature and the transition is suppressed by the Sr doping. All Nd$_{1-x}$Sr$_x$NiO$_3$ thin films show a metallic behaviour and a slight resistance upturn at low temperatures.

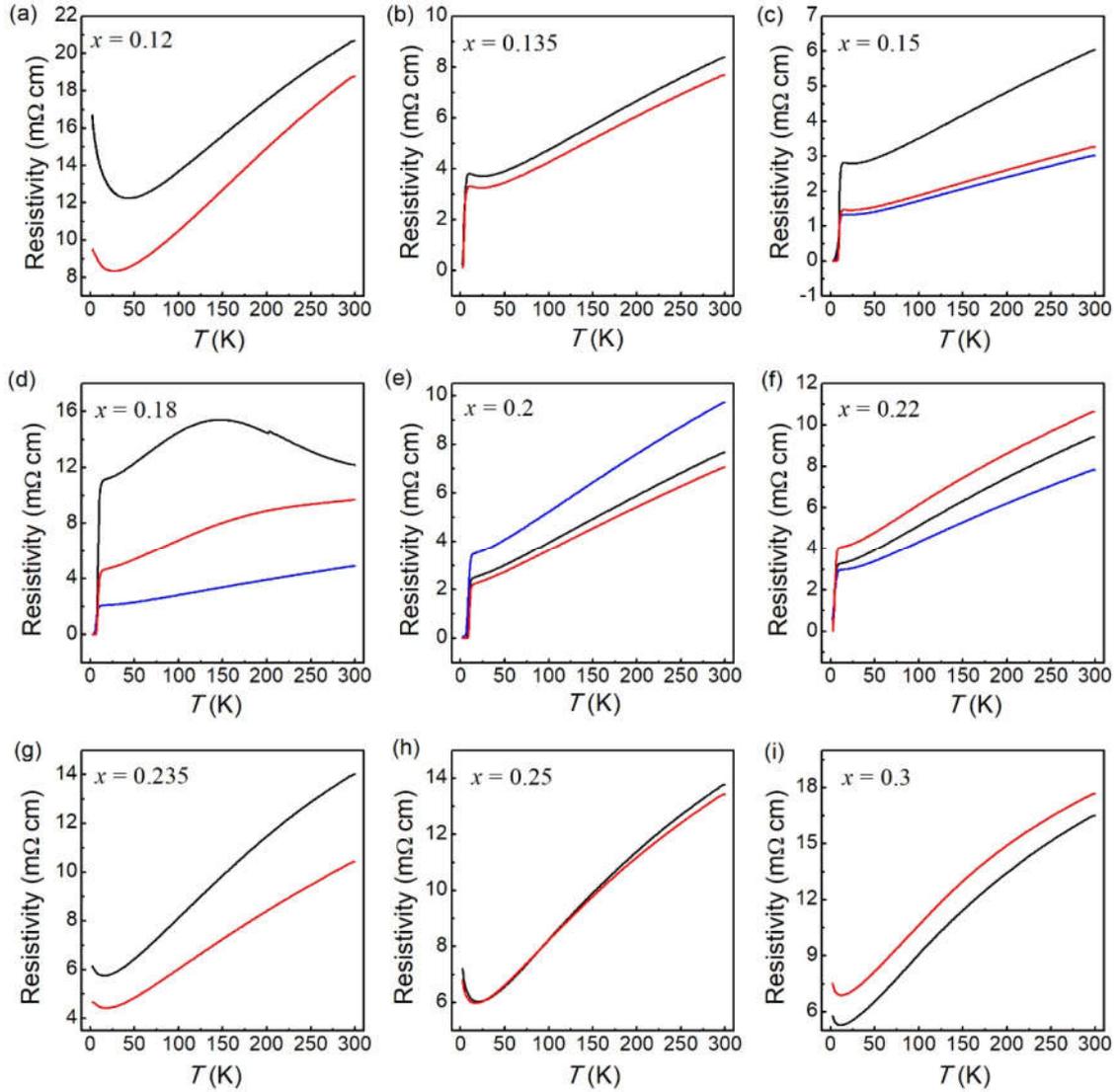

Fig. S5. The resistivity versus temperature ($\rho$-$T$) curves of various samples with $x$ from 0.12 to 0.3. (a) $x = 0.12$; (b) $x = 0.135$; (c) $x = 0.15$; (d) $x = 0.18$; (e) $x = 0.2$; (f) $x = 0.22$; (g) $x = 0.235$; (h) $x = 0.25$; (i) $x = 0.3$. The red curves for specific doping are the ones showed in Fig. 2 in the main text. The curves shown in other colours correspond to different samples not shown in Fig. 2 in the main text.

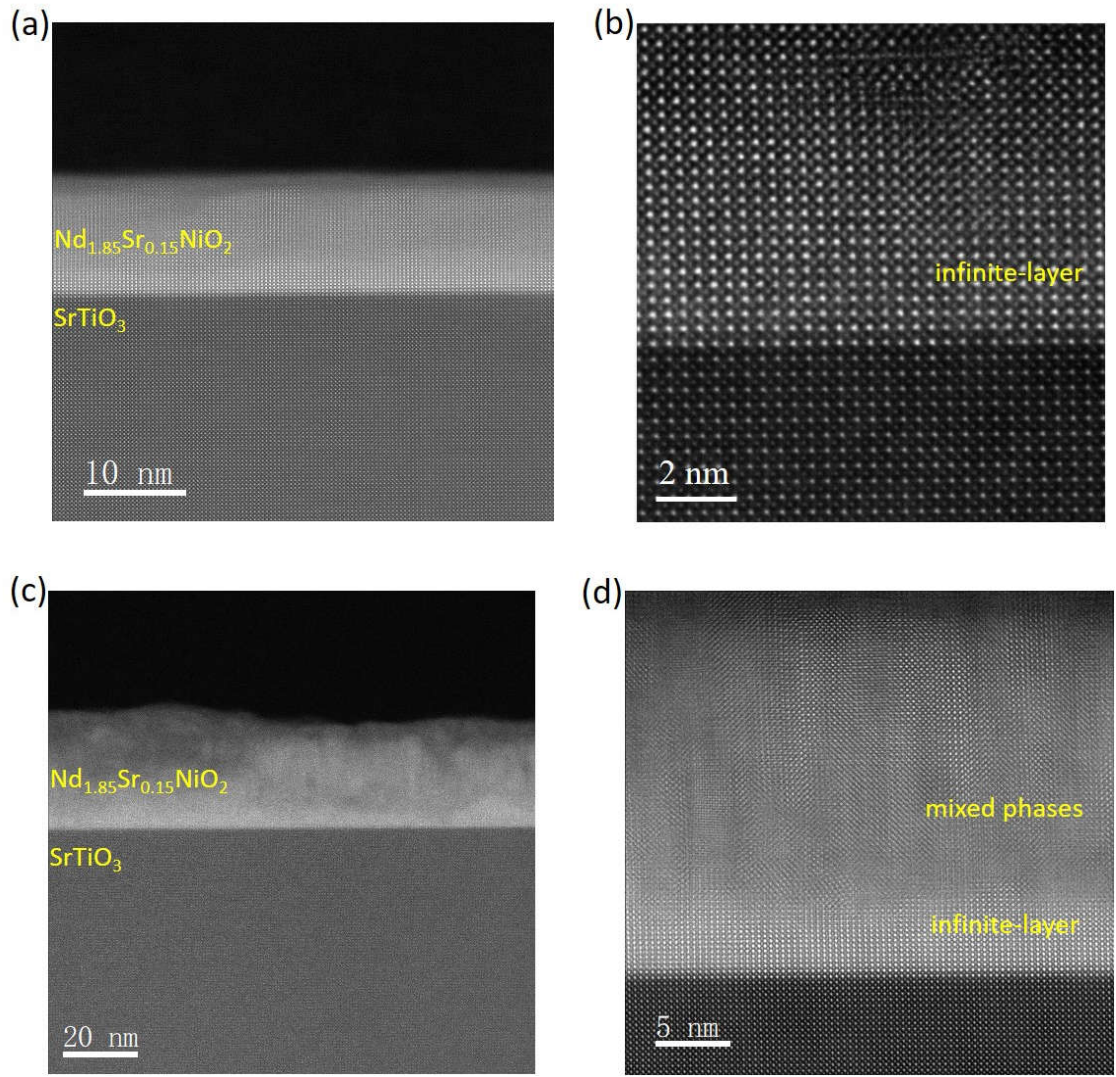

Fig. S6. (a) The HAADF-STEM image and (b) the magnified view of the ~10-nm $Nd_{0.85}Sr_{0.15}NiO_2$ on STO. (c) The HAADF-STEM image and (d) the magnified view of the ~35-nm $Nd_{0.85}Sr_{0.15}NiO_2$ on STO. For the 10-nm film, a clear infinite-layer structure is observed with no obvious defect throughout the layer. For the 35-nm film, the infinite-layer structure can also be seen at the bottom section of the film near the substrate. At the top section, there exists a mix between the infinite-layer and the Ruddlesden–Popper-type secondary phase. The 10-nm films show a similar superconducting transition and $R_H$ behavior to those for the 35-nm film as shown in Fig. S7.

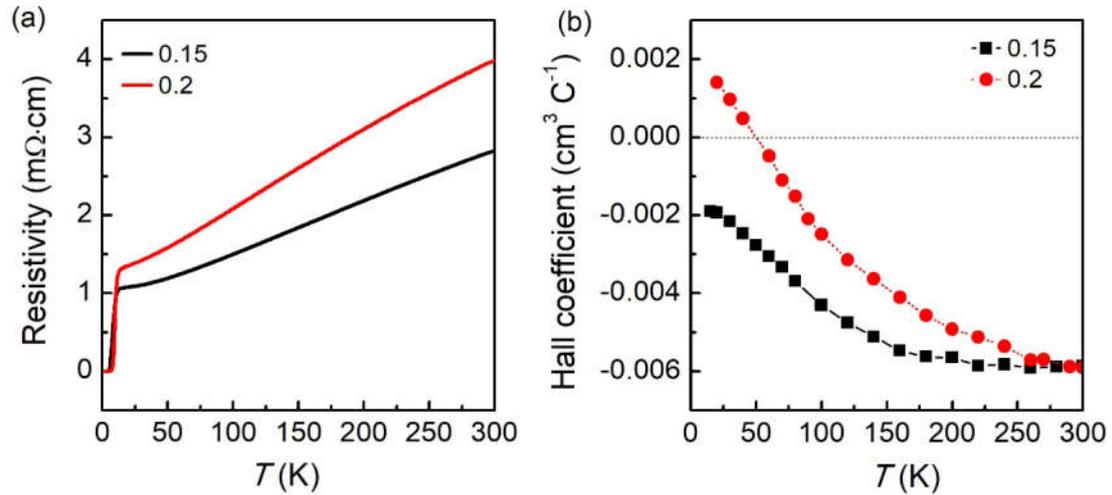

Fig. S7. (a) The resistivity versus temperature ($\rho$-$T$) curves of the ~10-nm thin films with $x = 0.15$ and $x = 0.2$. (b) The temperature dependence of the Hall coefficient $R_H$. The 10-nm films show a similar superconducting transition and $R_H$ behavior to those for the 35-nm film shown in Figs. 2 and 3 in the main text.

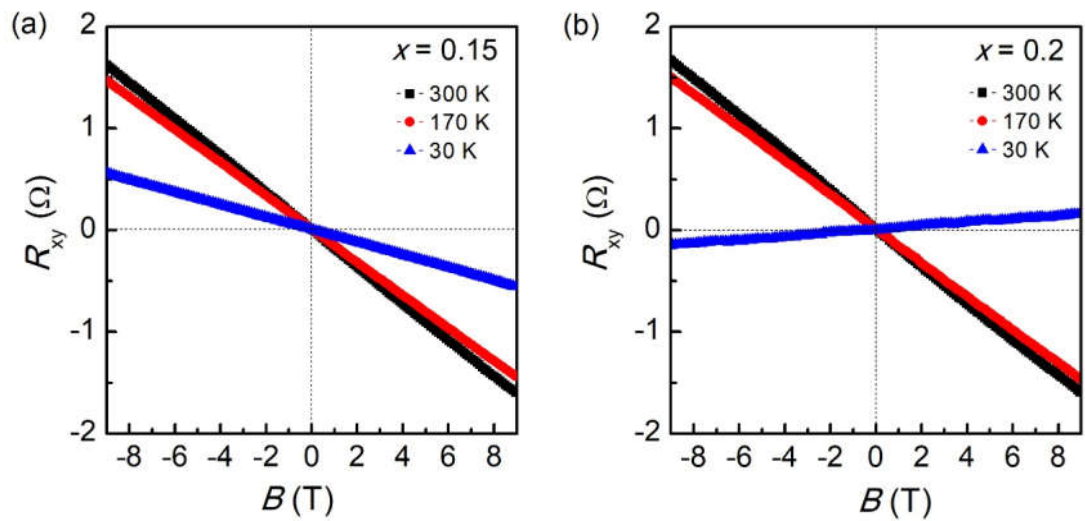

Fig. S8. The Hall resistances $R_{xy}$ of the thin films with two representative doping levels of (a) $x = 0.15$ and (b) $x = 0.2$. One can see that all the curves show a linear behavior.